\documentclass[12pt]{article}
\usepackage{cmpj2e}
\usepackage{graphicx}
\usepackage{amssymb}
\usepackage{epstopdf}
\def\be{\begin{equation}}
\def\ee{\end{equation}}
\def\bear{\begin{eqnarray}}
\def\eear{\end{eqnarray}}
\title{Towards an analytical theory for
 charged hard spheres }
\author{ Lesser Blum\refaddr{rutgers}, Domingo V. Perez Veloz\refaddr{PuertoRico}}
\addresses{
\addr{rutgers} Department of Mathematics, Hill Center, Busch Campus, Rutgers University, Piscataway, N.J. 08854
\addr{PuertoRico} Department of Physics, Faculty of Natural Sciences, University of Puerto Rico, Rio Piedras, Puerto Rico 00931
}
\begin{document}
\maketitle
\begin{abstract} 
Ion mixtures require an exclusion core to avoid collapse. The Debye Hueckel theory, where ions are point charges, is accurate only in the limit of infinite dilution. The MSA is the embedding of hard cores into DH, and is valid  for higher densities. The properties of any ionic mixture can be represented by the single screening parameter $\Gamma$ which for the equal ionic size restricted model is  obtained from the Debye parameter $\kappa$. This $\Gamma$ representation (BIMSA) is also valid for complex / associating  systems, such  as the general n-polyelectrolytes. The  BIMSA is the only theory that satisfies the infinite dilution limit of the DH theory for any  chain length.  Furthermore, the contact pair distribution function calculated from our theory agrees with the Monte Carlo of Bresme ea.(Phys. Rev. E {\textbf 51} 289 (1995)). 
\end{abstract}
\section{Introduction}
Stable ionic mixtures require an exclusion core to avoid the collapse of the system. For this reason the Debye Hueckel (DH) \cite {dh23}theory, in which the ions are point particles, is accurate only in the limit of infinite dilution. The MSA \cite{py,lebperc,waleb} that is an embedding of the DH theory into the hard-core Ornstein Zernike (OZ) equation is valid  for high densities, and is even asymptotically exact at infinite density \cite{yrlb86}. In the MSA the properties of any ionic mixture can be represented by a single screening parameter 
$\Gamma$ \cite {lb75} which in the simplest equal ionic size restricted model is  obtained from the Debye screening parameter $\kappa$: 
\bear
{ \Gamma}={1\over 2 \sigma}\{\sqrt{1+2 \kappa\sigma}-1\}
\eear
$\kappa$ is the inverse of the Debye screening length defined by
\be
\kappa^2 = \frac{4 \pi \beta e^2}{\varepsilon} \sum_k \rho_k z_k^2
\label{5.4}
\ee
where $z_k$ is the electrovalence, $\rho_k$ is the density of component k, $\beta$ is the Boltzmann thermal factor, $\sigma_k$ is the hard core diameter, and $\epsilon$ is the dielectric constant.
The one parameter representation \cite{enbl05} is valid for a number of complex and associating  systems such as dimers, and even polymers, where
 as in the Debye-Hueckel theory, the thermodynamic properties depend on a single screening parameter $\Gamma$: for the dimerization of equally charged hard ions we get
\bear
\kappa^2 {\alpha+\Gamma \sigma \over 1+ \Gamma \sigma}=4 \Gamma^2(1+\Gamma \sigma)^2
\label{eq5.4}
\eear
After a detailed analysis\cite{blube04,bebl96,bebl00} we can generalize  eq.(\ref{eq5.4})in the form
\bear
\kappa^2 {\mathcal F}(n,\alpha)=4 \Gamma^2(1+\Gamma \sigma)^2
\eear
For dimers
\bear
 {\mathcal F}(2,\alpha)={\alpha+\Gamma \sigma \over 1+ \Gamma \sigma}
\eear
In the limit of total association $\alpha=0$, and we recover the DH limiting law with the charge of the polyelectrolyte, $for \,\, any$  $n\ge 3$. Interestingly this will $ only $  happen if the correct virial (with the bridge diagrams) of the 3 body is used, and we get for linear homopolymer chains. Then after a lengthy calculation
\bear
\lim_{\rho \rightarrow 0}{\mathcal F}(n,0)\sim n^2;\quad correct\,\, DH \,\,limit\nonumber \\
\lim_{\rho \rightarrow \infty}{\mathcal F}(n,0)\sim n;\quad \,\, high \,\, density \,\,limit\nonumber \\
\label{eq4}
\eear
The criticality of ionic systems, initiated by M.E. Fisher  and collaborators is still a subject of current interest. \cite{fl93,aqua05,mefazp}.
In the original discussion of this problem Fisher and Levin \cite{fl93} used the DH \cite{dh23} theory in combination with  Bjerrum association \cite{bj26}. It is clear  that to get a mathematically well defined system $ all$ the ions need to have an exclusion core. Most of the subsequent work is directed at the inclusion of the excluded core in the nonlinear Coulomb and the association problem.  In our own previous work \cite{jiagbl02} we used different combinations of the binding mean spherical approximation (BIMSA) \cite{bebl96,bebl00} and various treatments of  ion association, all derived  from the work of Bjerrum \cite{bj26}. The best agreement with computer simulations was obtained from the SIS-BIMSA treatment of the association constant (Jiang, Blum, Bernard, Prausnitz and Sandler \cite{jiagbl02}. In the SIS approximation of Stell et al \cite{stelletal} the chemical association constant is computed from the contact pair correlation function.
A  discussion of the merits of the different approximations was recently given by  Aqua, Banerjee and Fisher \cite {aqua05}). The real problem here is that, as it has been shown by Wertheim for the associating systems, the classical virial expansion does not converge \cite{msw84,msw88} and a new MSA (the BIMSA), based on the Wertheim-Ornstein-Zernike equation (WOZ) has to be formulated. As has been shown elsewhere \cite{bebl96,bebl00}, for dense systems, the scaling solution and the remarkably simple thermodynamics  of the MSA applies to the new theories but with a renormalized screening constant, $\Gamma^N$. In the ESMSA the low density limits are also included \cite{bla06}.
The real issue is that we need an  internally consistent procedure and therefore there is an open (big) question about their reliability and accuracy that we will try to answer.\\

 \section {Theory}
We wish a very happy birthday to Fumio Hirata on the occasion of his birthday\\
Ever since the DH \cite {dh23} theory was formulated, almost a century ago, there has been a steady effort in ways to improve and extend the range of its multiple applications. Most of the theoretical effort has been dedicated towards the inclusion of the hard exclusion core and the nonlinearity of the Poisson-Boltzmann equation. We know now that \cite{yrlb86} that the MSA is asymptotically exact at very high densities. At lower densities the association limits yield exact conditions that are satisfied by closures of the Wertheim-Ornstein-Zernike equation: the exact DH limit at infinite dilution must contain the charges of the associated ions.
 The analytic solution of the MSA  As has been pointed out in this work since the parameter space of these systems is large, we need to have an accurate, fully analytical theory for the properties of these systems. The general problem has been discussed in the past by several authors ( see for example  \cite{yka97}). In recent work we have proposed a new theory, the ESMSA which could be such a theory, since there is only one screening  parameter $\Gamma$, a in the MSA, but also the degree of dissociation, ionic diameters, effective dielectric constants enter in the calculation through the contact pair correlation function. 

The most interesting feature of the MSA even for complex systems, is that, as in the Debye-Hueckel theory, the thermodynamic properties depend on a single screening (scaling) parameter $\Gamma$:
\be
4 [\Gamma]^2 (1+ \Gamma\sigma)^2 = \kappa^2 \frac{(\alpha +\Gamma
\sigma)}{(1+\Gamma \sigma)}
\label{5.2}
\ee
$\kappa$ is the inverse of the Debye screening length defined by
\be
\kappa^2 = \frac{4 \pi \beta e^2}{\varepsilon} \sum_k \rho_k z_k^2
\label{5.4a}
\ee
and $\alpha$ is the degree of {\it dissociation}. The remarkable property of this equation is that it yields the correct asymptotic limits for zero density and also for very large density \cite{yrlb86}, where
\bear
{ \Gamma}={1\over 2 \sigma}\{\sqrt{1+2 \kappa\sigma}-1\}\sim \left\{\sqrt{ \kappa\over 2\sigma}\right\}
\eear
 In the case of associating (polymerizing) ions the BIMSA for the restricted primitive model
yields a very simple expression  $\Gamma^B$
This result applies to Polyelectrolytes and explains the remarkable agreement with the simulations of Orkulas et al. \cite{okp03}\\
The question is which is the best interpolation scheme between these two regimes?
One possible answer can be obtained by testing different combinations as was done in our previous work \cite{jiagbl02}:
However it is clear that one cannot separate the sources of errors in the analysis of the criticality. However a more illuminating perspective can be obtained considering the simulations of Bresme et al \cite {bresm95}, as we will see it below.
The usefulness of the MSA resides in the fact that the scaling solution is formally valid for the general mixture of arbitrary charge and size ions.

It has been shown by Rosenfeld and Blum (Y. Rosenfeld and L. Blum,\cite{yrlb86} that the MSA is asymptotically exact at very large density. But the point remains that at low temperature ${\mathcal ANY}$ theory based on the normal Ornstein Zernike (OZ) equation will fail, because 
it will not be able to reach the full association limits. This is corrected by the Wertheim Ornstein Zernike (WOZ)\cite{msw88} equation, which as has been shown to work remarkably well even for the limit of infinitely long polyelectrolyte chains\cite{blube04}.\\

For many applications it is important to have an analytical, but at the same time an accurate theory of electrolytes. This is not an easy task, but recent advances have made this an attainable goal. One important ingredient of this theory is presumably the contact pair correlation function (PCF). In this communication we evaluate a simple functional form of the contact PCF, using the Monte Carlo simulations of Bresme et al. \cite{bresm95}. We find that the best analytical representation is obtained from this formula. It is increasingly erroneous at high charges. One possible reason for this failure is the fact that the simulations do not satisfy the PST, as is suggested by some preliminary simulation runs. Another source of errors is the neglect of the field parameter $\eta$.

\section{The contact pair distribution function}
In the present contribution we show that the contact pair distribution function for a hard ions mixture can be represented explicitly by a functional of $\Gamma$ 
 \cite{bebl96,bebl00,bla06}:
\bear
 g_{ij}^{00}(\sigma_{ij})
= g_{ij}^{HS}(\sigma_{ij})e^{
-  \frac{\beta e^2}{\sigma_{ij}\varepsilon_0} X_i^{01} X_j^{01}}
;\quad
 X_i^{01}={z_i-\eta \sigma^2_i\over 1+\Gamma \sigma_i}\simeq{z_i\over 1+\Gamma \sigma_i}
\label{eq10}
\nonumber\\
\eear
$\eta$ is a mean field parameter which depends on  many body interactions, such as $\alpha$, usually very small, and  $\epsilon$ is  an  effective dielectric constant. 
The hard core contact correlation is from the Carnahan-Starling approximation 
\cite{cs69}
\bear
g_{ij}^{HS}(\sigma)={1-{\pi\over 12} \sum_i \rho_i\sigma^3\over(1-{\pi\over 6} \sum_i \rho_i\sigma^3)^3}
\eear
but in our case can be taken to be 1, since our system is very dilute.\\
Indeed the proper discussion of this system is to treat it as a mixture of ions and dipoles for the associated part\cite{lb02}.
We get for our restricted equal diameter problem
\be
\sigma_{ij}=\sigma_i=\sigma_j=\sigma 
\ee
The simplest approximation is to take $\eta=0$, which means that we are neglecting higher order correlation effects\cite {bebl00}. Then from eq.(\ref{eq10}) we get
\bear
 g_{ij}^{00}(\sigma)
= e^{
- {z_i z_j\beta e^2\over\varepsilon\sigma (1+\Gamma \sigma)^2}}
\label{eq12}
\eear
The results of the calculation are shown in figure 1. The agreement for $\beta e^2\over\sigma$ up to 16 is quite good, and much better than  the HNC theory \cite{chap94} which does not easily converge or the INV-C of Duh and Haymet \cite{dh92} which is shown in fig.1.
However in our theory the error in the last point is sizable, and we have no explanation for this at the time, other than the suspicion that ths may be a consequence of the failure to satisfy the PST \cite{pst82}. 
\begin{figure}[t]
\begin{center}
\resizebox{3in}{!}{\includegraphics{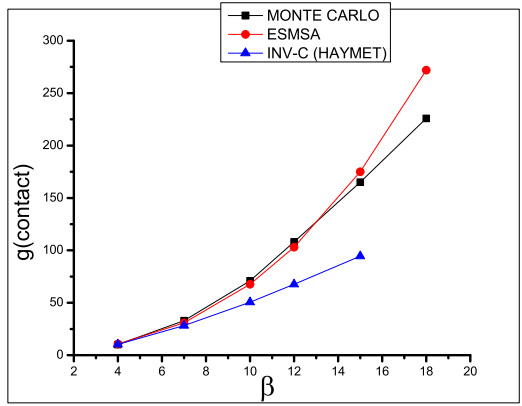}}
\caption{Comparison of the contact pair distribution function for $g^{ESMSA-BIMSA}_{+-}$.
The squares are the Monte Carlo results of Bresme et al.\cite{bresm95}. The triangles are the INV-C theory of Haymet et al \cite{dh92}. The circles are the ESMSA-BIMSA with $\eta=0$ and an effective  dielectric constant of $\epsilon=1.17$}(which also depends on the association constant $\alpha$)
\label{fig:energy1}
\end{center}
\end{figure}
\section*{Acknowledgements}
The authors are indebted to the NSF-PREM program for finantial support.
We wish a very happy 60th birthday to Prof. Fumio Hirata.


\begin{thebibliography}{99}
\bibitem {dh23} P. Debye and E. Hueckel, Phys. Z. (1923),\textbf{24}, 183 . 
\bibitem{py}  Percus J.K. and Yevick G.,  Phys. Rev.,1964, ,{\textbf 110} 250 .
\bibitem{lebperc}J. L. Lebowitz and J. K. Percus, Phys. Rev.  (1966),{ \bf 144} 251  . 
\bibitem{waleb} E. Waisman and J. L. Lebowitz,
 J. Chem. Phys.  (1970),{\textbf 52} 4307 .
\bibitem{yrlb86}  Rosenfeld, Y.  and  Blum, L.,
{J.  Chem.  Phys. }, ( 1986) , {\textbf 85} 1556. 
\bibitem{lb75}L. Blum, Mol. Phys. (1975),{\textbf 30} 1529  . 
\bibitem{enbl05} A. Enriquez, L. Blum,    Mol. Phys. ,(2005), {\textbf 103}, 3201 .
\bibitem{blube04}  O.Bernard and L. Blum, { Proc. International School        of Physics "Enrico Fermi", Course CLV,} Mallamace, F. and  Stanley, H. E., Editors, IOS Press, Amsterdam, (2004), {\textbf 155}, 335 .
\bibitem{bebl96} O.Bernard and L. Blum, J. Chem. Phys. (1996) {\textbf 104}, 4746).
\bibitem{bebl00} O.Bernard and L. Blum, J. Chem. Phys. ,(2000), {\textbf 112}, 7227)
\bibitem{fl93}M. E. Fisher, Y. Levin, Phys. Rev. Lett. (1993), \textbf{71}, 3826 . 
\bibitem{aqua05} J-N Aqua, S. Banerjee, M. E. Fisher,  Phys. Rev.(2005), E \textbf{72}, 041501:1-25.
\bibitem{mefazp} Y. C. Kim, M. E. Fisher,  A.Z. Panagiotopoulos, Phys. Rev.(2005),\textbf{95}, 195703:1-4.

\bibitem{bj26} N. Bjerrum, Kgl. Dan. Vidensk. Selsk. Mat.-Fys. Medd. (1926),{\textbf 7}, 1 .

\bibitem{jiagbl02} J. W. Jiang , L. Blum, O. Bernard, J. M. Prausnitz and S. I. Sandler, J. Chem. Phys.(2002),{\textbf 116}, 7977 .
\bibitem{stelletal} G. Stell et al.  J. Chem. Phys.(1989), {\textbf 91}, 3618 ,  J.  Phys. Chem.,(1996),{\textbf 100}, 1415 , J. Chem. Phys.,(1995),{\textbf 102}, 5785 .
\bibitem{msw84} M.S.Wertheim, J. Stat. Phys.  (1984), {\textbf 35}
19:35    ,
 (1984),{\textbf 42l 459: 477 .
\bibitem{msw88}M.S.Wertheim, J. Chem. Phys.,(1985), {\textbf 85} 2929 ,
 (1987),{\textbf 87} 7323   , (1988),{\textbf 88} 1214   .
\bibitem{yka97} Yu.  V.  Kalyuzhnyi, L.  Blum, M.  F.  Holovko and I.  A.
Protsykevytch,   
Physica (1997), A {\textbf 236} 85 .
\bibitem{okp03} G. Orkulas, S. Kumar , A.Z. Panagiotopoulos, Phys. Rev. Letters (2003), \textbf{90}, 048303-25.
\bibitem {bla06} L.Blum , M. Arias, Mol. Phys.,(2006), {\textbf 104}, 3801 . 
\bibitem{bresm95}F. Bresme, E. Lomba, J.J. Weis, J.L. F. Abascal, Phys. Rev. E (1995),{\textbf 51}, 289 .
\bibitem{pst82}  L. Blum,  Ch. Gruber,   J. L. Lebowitz  and  Ph. A. Martin,  
Phys. Revs. Letters, (1982), {\textbf 48},  1769 . 

\bibitem{cs69} N.F. Carnahan, K.E. Starling,  J. Chem. Phys (1969),{\textbf 51 }, 635 . 
\bibitem{dh92}  D.M.Duh and A.D.J. Haymet { J. Chem. Phys.},(1992), {\textbf 97 },7716 .
\bibitem{chap94} M. Llano-Restrepo and W.G. Chapman, { J. Chem. Phys.} (1992),{\textbf 97 }, 2046 , (1994),{ \textbf 100 },5139 .
\bibitem{lb02} L. Blum  J. Chem. Phys.}
 {\textbf 117 }, 756 (2002). 
\end{thebibliography}
\end{document}